# Size and strength of self-excited dynamos in Jupiter-like extrasolar planets

Mohamed Zaghoo[1,2] & G. W. Collins[1,3]

1-Laboratory for Laser Energetics, University of Rochester, NY, USA
2-Lyman Laboratory of Physics, Harvard University, Cambridge, MA, USA
3-Department of Physics & Astronomy and Mechanical Engineering, University of Rochester, NY, USA

**The magnetization of solar and extrasolar gas giants is critically dependent on electronic and mass transport coefficients of their convective fluid interiors. We analyze recent laboratory experimental results on metallic hydrogen to derive a new conductivity profile for the Jovian-like planets. We combine this revised conductivity with a polytropic-based thermodynamic equation of state to study the dynamo action in 100 extrasolar giant planets varying from synchronous hot jupiters to fast rotators, with masses ranging from $0.3M_J$ to $15M_J$. We find conducting cores larger than previous estimates, but consistent with the results from Juno, suggesting that the dynamos in the more massive planets might be shallow-seated. Our results reveal that most extrasolar giants are expected to possess dipole surface magnetic fields in the range of 0.1-10 Gauss. Assuming radio emission processes similar to our solar giants, most characterized planets should emit radiation with a maximum cyclotron frequency between few and 30 MHz, lower than previous estimates. Our work places new bounds on the observational detectability of extrasolar magnetic fields.**

The interiors of solar gas giants are seats for large-intrinsic magnetic fields that are the result of their convective-driven dynamos, whereas their surfaces feature sizable zonal flows. These flows are bound to interact with the magnetic field at depths that could extend to as deep as the highly conducting regions (Stevenson 1983). In extrasolar giants, strong magnetism is expected to play an important role in the thermal evolution, mass history, and atmospheric properties of these planets. For example, the interaction of planetary magnetic fields with the weakly ionized atmospheres of hot giants have two major consequences. First, it can result in considerable kinematic induction ohmically depositing additional heat in the interiors and or atmospheres. Such phenomenon is thought to account for the inflated radii of several transiting hot Jupiters, the so-called radius anomaly (Batygin & Stevenson 2010; Perna et al. 2010). Second, the outflow of these atmospheric plasma winds can lead to the formation of a current-carrying magnetodisk, that dominate the planetary megnetospheres (Trammell et al. 2011). Such magnetodisks would dictate the rate of particle loss in the atmospheres especially in strongly insolated extrasolar giants. Unsurprisingly, the importance of these magnetic effects on the giants' atmospheres and interiors is critically dependent on the surface field strength. (Menou 2012) showed that as the magnetic field strength increases, so does the effective equilibrium planetary temperature. Accordingly, it is possible to describe the relation between in the transiting giants' radii in relation to their effective temperature by the variations in their surface poloidal fields.

Understanding these magnetization effects in extrasolar giants necessitates better characterization of the size and the strength of their respective planetary magnetic dynamos. To date, direct detection of the emission resultant from the magnetization of extrasolar planets have proved elusive (Bastian et al. 2000; Grießmeier 2015; Joseph et al. 2004; Sirothia et al. 2014). With little observational guidance, it is important to examine the dynamo action in Jupiter and apply the results to the extrasolars giants. Two developments in this area are revising previous constraints on the size and strength of Jovian-like dynamos. First, new experimental data on the conductive properties of Jupiter's liquid metallic core show the conductivity of fluid hydrogen at the corresponding depths are larger than currently assumed from previous shock-wave experiments or predicted in ab-initio based conductivity profiles (Zaghoo & Silvera 2017). Secondly, the first few perijoves from Juno space mission reveal a striking spatial variation of the magnetic field at radial distances much closer to the Jovian surface (Bolton et al. 2017). The field's strength measured magnitude, 7.6 G, is larger than what most pre-Juno magnetization models predict, 4.3 G (Connerney et al. 2017). Thus, both observational and laboratory experimental results strongly suggest that the dynamo is operating at much shallower depths than hitherto assumed.

This paper updates and extends previous reports describing the strength of the intrinsic dipole magnetic fields of extrasolar gas giants and their associated magnetospheres. Early studies (Cuntz et al. 2000; Farrell et al. 1999) scaled the Jovian magnetic values for extrasolars planets while assuming similar internal structure to that of Jupiter. Sanchez 2004, Lazio et al. 2004, Grießmeier et al. 2004 presented the results of multiple scaling laws for different sets of selected exoplanets, assuming constant core sizes. More recent results studying magnetic evolution of hot Jupiters and their associated radio emission flux used an internal heat-flux based scaling law and found magnetic fields strengths that are an order of magnitude larger than previous studies(Reiners & Christensen 2010). These results were reanalyzed to include effects of deposited stellar radiation fluxes suggesting even significantly higher surface magnetic fields (Yadav & Thorngren 2017). Here, we present a new experimentally based internal conductivity that could better account for the Juno's observation. We combine it with a polytropic equation of state to better estimate the size of dynamo core in 100 extrasolar gas giants. Estimates of their surface magnetic field strength, magnetic dipolar momentum and radio emission frequencies are also presented in comparison to several scaling laws and previous studies. We discuss the implications of these results on the limits for future low frequency ground based and space observations.

**Conductivity profile**

Fluid hydrogen constitutes almost 90% of Jupiter-like planetary mass and the conditions along their isentrope permit states that are molecular, partially ionized (semiconducting) and metallic. The presence of 9% helium is arguably less important due to its high ionization energy, even at the internal conditions of interest. Near the surface, the electrical conductivity is dominated by traces of alkali metals low ionization, mostly due to stellar irradiation and collisions with highly energetic particles. At such low densities, hydrogen exists as a strongly bound molecular fluid with a bandgap of 15.5 eV and conductivity eventually takes on the dilute gas values. As gravitational pressure increases, many-body physics effects are expected to lower the ionization energy and progressively its electronic band gap, enabling some conduction by thermal activation of carriers. In this region (0.2-0.7 g/cc), most MHD simulations approximate the conductivity as that of a semiconductor, where the fractional occupancy of conducting states is proportional to $\exp(-E_g/2kT)$ with k being the Boltzmann constant, $E_g$ is the bandgap and T is the temperature. At sufficient densities and temperatures, the average energy of electrons becomes larger than the hydrogen's ionization potential, and the system pressure-ionizes into a degenerate metallic state (Hubbard et al. 1997). The conductivity of this state is typically proportional to the carrier density, $n_e$ which is related to the mass density $\rho$, via and $n_e = (2\,A\,\rho)/M$, where $A_v$ is the Avogadro's number and M is the molecular weight of hydrogen.

Recently, laboratory experiments utilizing both static compression and precompressed laser driven shockwave techniques probed metallic hydrogen optical conductivity over a wide range of pressures spanning 20-500 GPa (Brygoo et al. 2015; Dias & Silvera 2017; Zaghoo & Silvera 2017). The optical response as a function of energy was used to determine hydrogen's static electrical conductivity in the free-electron model. The results indicate that metallic hydrogen conductivity at conditions corresponding to 0.86 Rj is 6-8 times larger than previously reported (Zaghoo & Silvera 2017). In Fig 1, we plot the experimental values for the fluid metallic state from the static and shockwave optical experiments. We note that optical data collected at 500 GPa was measured at cryogenic temperatures. Its corresponding fluid state, above the melting point (~600-1000 K), should have lower conductivity, since the electron transport collisional frequency scales inversely with temperature. We apply the Mott theory for liquid conduction to determine the reduction in conductivity close to the melting point(Mott 1972), and assume that the conductivity once in the fluid state doesn't feature a steep temperature, as found in other liquid metals. To constrain our fit in the low-pressure region, we have included the shockwave electrical resistivity datum at 22 GPa and 4400K (Nellis et al. 1992). Our proposed profile is compared to that obtained from Density Functional Theory (DFT) ab-initio calculations (French et al. 2012), widely in use as an input for dynamo simulations. As shown in Fig 1, the conductivity rises precipitously below $R_J$=0.9 and becomes shallow above 0.88$R_J$. The ab-initio theoretical curve is revealed to underestimate conduction in the metallic regimes. To further illustrate this, we compare the ab-initio values to those expected for fluid metals within the Ioffe-Regal Mott minimum metallic conductivity limit(Mott 1972). This limit is important in the theory of disordered conductors and doesn't feature temperature dependence.

$$\sigma_{MIR}(r) = e^2/4\hbar \sqrt[3]{2\,A_v\,\rho(r)/M} \quad (1)$$

As shown in Fig.1, the ab-initio calculation falls short of the minimum metallic criterion, even at conditions where hydrogen is experimentally shown to assume the metallic state. Interestingly, the conductivity profile shows a steep dependence on density that is not captured by the theoretical models or the approximation employed in recent anelastic studies of the dynamo.

**Onset of the Dynamo action**

The appropriate measure for the feasibility of the dynamo action, and correspondingly the depth of its field generation region, is the magnetic Reynolds number, $R_m = v_{\text{conv}} L/\lambda(r)$. Here, L(r) is the relevant conductivity scale height, $\lambda(r)$ is the magnetic diffusivity profile (inverse of the electrical conductivity), and $v_{\text{conv}}$ is the characteristic azimuthal flow velocity. L(r) is, by definition, much smaller than the planetary radius, and its typical value near the conductive region is taken to be ~$10^3$ km. The majority of dynamo simulations prescribe a magnetic Reynolds number 10-100 as a requisite to sustain the dynamo action. Here, we take $R_m = 50$ as an average value. In order to determine this metric, one needs to calculate the convective flow velocity. (Christensen & Aubert 2006) suggested an internal velocity of $2\times10^{-2}$ ms$^{-1}$, based on extensive suite of numerical studies of the relevant parameter space. This number is consistent with studies of the time-varying field of Jupiter over a span of 44 years, which reveal an internal flow velocity of $10^{-2}$ ms$^{-1}$. Alternatively, another bound on the flow velocity could be determined from the internal planetary heat flux. The mixing length theory (MLT) or the magnetospheric MAC balance are both acceptable framework to calculate these flow velocities. Both are reasonably justified in regions representative of the ones of interest: as the former relies on persistence of large convection motions, while the latter assumes that in the field generation region, the buoyancy, Lorentz and Coriolis forces are comparable. We focus here on the mixing length theory.

$$v_{conv\ (MLT)} = \left(\frac{\alpha\delta}{4}\frac{P}{\rho T C_P}\frac{F_{conv}}{\rho}\right)^{\frac{1}{3}} \quad (2)$$

Here $F_{conv}$ is the convective flux, $\alpha$ is the mixing length parameter and is close to unity, $\delta = (\partial \ln\rho/\partial \ln T)_P$ is usually taken to be independent of depth, $C_p$ is the specific heat at constant pressure. The convective velocity is thus proportional to $(\frac{F_{conv}}{\rho})^{1/3}$. For superadiabatic conditions, typical of deep levels of gaseous planets, the radiative gradient is much larger than the temperature gradient owing to the high conductive opacities of hydrogen. Thus, the convective flux is thus equal to the internal heat flux $F_{int}$.

For Jupiter, the internal heat flux, $F_{int\ J} = 4.5$ W/m$^{-2}$, corresponding to a convective flow velocity of $4\times10^{-2}$ ms$^{-1}$. We note that smaller velocities, of $10^{-3}$ ms$^{-1}$, could be found from the MAC balance. The corresponding Magnetic Reynolds number calculated from the conductivity profile and MLT estimates is shown in the inset of Fig. 1. The higher electrical conductivity indicates that conditions favorable for the dynamo action could develop at depths even shallower than 0.91 Rj (Rm~100). It is noteworthy to highlight that this depth is in accord with the estimates found in the most recent study of the physical parameters space of the dynamo action in Jupiter (Duarte et al. 2018). The density corresponding to this depth is also where most ab-initio calculations find that fluid hydrogen transition to a degenerate atomic-like state, consistent with our suggested onset for the dynamo action. Using the same MLT assumption in extrasolar planets, the convective flow could be calculated provided knowledge of the internal heat. This heat flux is a complex function of mass, and age, however, evolutionary models developed by (Burrows et al. 2001; Burrows et al. 1997) provide a reasonable framework. In these models, the convective heat flux at the outer planetary boundary could be obtained by diving the net planetary time-dependent luminosity, $L_p$ by the size of the metallic core, $R_c$.

$L_p$ could be described by $\sim 6.3 \times 10^{23}\ erg \left(\frac{t}{4.5\ Gyr}\right)^{-1/3} \left(\frac{M_p}{M_J}\right)^{2.64}$ (3)

and thus

$$F_{int} = F_{jnt\ J} \left(\frac{t}{4.5\ Gyr}\right)^{-\frac{1}{3}} \left(\frac{M_p}{M_J}\right)^{2.64} \left(\frac{R_c}{R_{cJ}}\right)^{2.64} \quad (4)$$

The average core densities were estimated by assuming $\rho \ \alpha\ 3M_P/4\pi R_P^3$, and a detailed calculation for the size of the dynamo cores, $R_c$, is provided below. For planets with masses of 0.3–15 $M_J$ and ages similar to our solar giants, 4.5 Gyr, $F_{int} \sim$ 0.3–1200 Wm$^{-2}$ is expected. The corresponding convective velocities are 0.002-0.1 ms$^{-1}$ (see Fig.2). Younger planets of 1 Gyr are thus expected to possess heat fluxes that are twice those of solar ages and accordingly slightly higher velocities. Our results are in good agreement with those estimated by (Sanchez-Lavega 2004) who studied planets with masses of 0.3-10 $M_j$ and 1-10 Gyr age. He determined that for orbital periods spanning slow (3-4 days) to fast rotators (2-5 hrs), convective velocities should be in the range of 0.006-0.6 ms$^{-1}$.

**Size of the conducting cores**

To estimate the size of the dynamo field generation core for extrasolar giants as a function of their respective masses, we use a Polytropic Equation of State (P EOS) to construct their pressure-density profiles. Such EOS assumes that the pressure, P, closely follows a density profile with a polytropic relation of $P(r) = K\rho^n(r)$. Here, n is the polytropic index and $K$ is a constant that is determined by the mass of the planet $M$ and the boundary condition that the density must vanish at the planet's surface $R_p$,
$K = 2GR_p^2/\pi$

The radial density profile is determined from the Lane-Emden equation

$$\frac{d}{dr}\left(r^2 \rho_0^{n-2} \frac{d\rho_0}{dr}\right) = -\frac{4\pi G}{nK} r^2 \rho_0 \quad (5)$$

A polytropic index of n=2, (Hubbard 1977; Hubbard & Stevenson 1984), permits a closed form solution for the density

$$\rho(r) = \frac{\pi M_p}{4R_p^3} \frac{\sin(\pi r/R_p)}{(\pi r/R_p)} \quad (6)$$

In Fig.3, we compare our calculated polytropic EOS to those determined from the more sophisticated ab-density functional initio models.

Recent laboratory experiments have furnished strong evidence for a sharp first-order phase transition between liquid molecular and the atomic metallic phase in the region of 100-150 GPa and temperatures of 1800-1000 K (Zaghoo et al. 2016; Zaghoo & Silvera 2017). However, the Jupiter-like adiabat features much higher temperatures (~ 4500K) than those found in those experiments at the same densities. The crossover from the molecular hydrogen to the metallic envelope is thus likely continuous around 50 Gpa or ~ 0.7 g/cc. Using this transition density in the relation above, yields a conducting core r=$R_c = 0.89 R_J$ for Jupiter, very close to the 0.91 value specified above. We have, thus, solved this density distribution relation analytically for the 100 extrasolar giants considered here to find the size of their respective conducting cores. The mass and radius of these planets were taken from the current census. To categorize our results in terms of evolutionary models, we adopt the results of (Baraffe et al. 2003) who studied the mass-radius relationship of extrasolar giants with ages of 1,5 and 10 Gyr. For planets with radii close to Rj and masses between 0.3-3Mj, we find that the size of conducting cores increase with planetary masses. However, the relationship is hardly linear, see Fig 4. This range of masses spans planets with age ~1-5 Gyr, with the more massive ones close to our solar age. Those massive and older giants feature dynamo-size cores in the order of 0.9-0.99 of their respective radii. On the other hand, for the same mass, an increase in the radius yields smaller dynamo core sizes. Our calculated dynamo core sizes are overall different from previous conducting cores estimates, which employed simple scaling relations that are either based only on mass (Curtis & Ness 1986) or both mass and radius (Grießmeier et al. 2004). We note that these previous results return unphysical values (rc>R) for more massive and hot jupiters (Grießmeier et al. 2005). Additionally, our results are also overall higher than those previously estimated in (Sanchez-Lavega 2004), where the transition pressure to the metallic envelope was assumed to take place around 200 GPa, and the planets' masses or radii was taken from (Burrows, et al. 1997) evolutionary models rather than the extrasolar census.

**Magnetic field strength & radio emission**

In self-excited dynamos regimes, the balance between the Lorentz and Coriolis forces allows an estimate for the strength of the generated field, the Elsasser approximation (Stevenson 1983). Dynamo generation occurs at an Elsasser number $\Lambda = B^2/\rho\omega\lambda\mu_0$~1, where $\rho$ is the characteristic transition density at the top of the dynamo region and $\omega$ is the planetary angular rotation rate.

$$B_{rms} = \sqrt{\rho\omega\lambda\mu_0} \quad (7)$$

For a mainly dipolar field, the strength of the field at the planetary surface is estimated by downward continuation from $r_c$, $B_{dip} = B_{rms}\, r_c^3$, and the dipolar magnetic moment is $M_E = B_{dip} R_p^3$.

For Jupiter, we find a surface dipole field strength of 7.1 G and a magnetic moment of 2.46x10$^{21}$ T.m$^3$. These results are in excellent agreement with the most recent magnetic data from the Juno space mission. Since the time of synchronization of most extrasolar giants is ~ few Myr (Showman & Guillot 2002), we have assumed that their rotational periods are equal to their orbital ones. We find that, with the exception of cvso-30b, most of them are expected to possess surface fields in the order of 1-8 G. The most intense magnetic moments are found to occur in young, and inflated extrasolar within the mass rage of 1-3 Mj. Unlike previous studies, our scaling results don't support the general idea that more massive giants should always possess stronger fields. However, our estimated magnetic moments scale with the product of the planetary mass and radius squared, $M_E \alpha M_P R_p^3$, see Fig 5. This scaling was applied successfully to our four solar giants, since their internal densities are all close to 1g/cc. Several other scaling laws have been derived based on known internal and observational parameters of solar planets(Busse 1976; Christensen 2010; Mizutani et al. 1992; Sano 1993; Stevenson 1983).

$$M_E \; \alpha \; \omega \varrho_c^{1/2} r_c^4 \quad (8) \qquad \text{(Busse 1976)}$$

$$M_E \; \alpha \; \omega \varrho_c^{1/2} r_c^3 \sigma_c^{-1/2} \quad (9) \qquad \text{(Stevenson 1983)}$$

$$M_E \; \alpha \; \omega^{3/4} \varrho_c^{1/2} r_c^{7/2} \sigma_c^{-1/4} \quad (10) \quad \text{(Mizutani, et al. 1992)}$$

$$M_E \; \alpha \; \omega \varrho_c^{1/2} r_c^{7/2} \quad (11) \qquad \text{(Sano 1993)}$$

$$M_E \; \alpha \; (\varrho \; F_{core}^2)^{1/3} r_P^3 \quad (12) \qquad \text{(Christensen 2010)}$$

With the notable exception of the analytical scaling law of (Christensen 2010), all the others depend on the internal properties of the conducting fluid as well as rotational period of these planets. Moreover, this particular scaling law predicts a dynamo core size of 0.83 in Jupiter as opposed to 0.9 currently inferred from Juno and laboratory experiments. Fig. 5 shows the magnetic moments results of the different scaling models, where $\sigma_c$ was taken from the values determined for Jupiter in Fig 1. We don't plot the moments calculated with (Christensen 2010) model, since it features a strong time-dependence. We find that most planets are expected to possess dipolar magnetic moments in the range of $1 \times 10^{20}$ -$6 \times 10^{21}$ T.m$^3$, and these bounds appear insensitive to the choice of the scaling model, although they differ per individual planet. Our results are consistent with the values found in recent spectroscopic observation and magnetohydrodynamic (MHD) studies of hot jupiters atmospheres. For example, our calculated magnetic moment for HD209458b and HAT-P-7 are ~ 0.12 and 1.9 $M_EJ$ respectively (see Table 1). These values are consistent with the 0.1 $M_EJ$ estimate found from atmospheric hydrogen atomic linewidth in HD209458b (Kislyakova et al. 2014), and the minimal field strength of ~6 G required to produce the observed wind variability in HAT-P-7 (Rogers 2017).

| Planet | $M_P$ (Mj) | $R_P$ (Rj) | $F_{conv}$ (4.5 Gyr) | $v_{conv}$ (m/s) | $M_E$ (T.m³) | $R_c$ | $B_{dip}$ (G) |
|---|---|---|---|---|---|---|---|
| τ-BOOTS | 3.87 | 1.3 | 87.85 | 0.046 | $1.67 \times 10^{20}$ | 0.93 | 2.9 |
| HAT-P-7 b | 1.74 | 1.43 | 11.7 | 0.07 | $4.67 \times 10^{20}$ | 0.81 | 5.2 |
| WASP-18b | 10.43 | 1.16 | 1358 | 0.03 | $6.28 \times 10^{20}$ | 0.98 | 6.4 |
| CVS0-30b | 6.2 | 1.91 | 159 | 0.11 | $1.93 \times 10^{21}$ | 0.88 | 6.6 |
| HD 209458b | 0.69 | 1.38 | 1.77 | 0.083 | $2.94 \times 10^{19}$ | 0.64 | 0.9 |
| OGLE-TR -109b | 14 | 0.9 | 4833 | 0.016 | $6.48 \times 10^{20}$ | 0.99 | 8.3 |

*Table 1: Calculated convective heat flux, flux velocity, magnetic moments, dynamo core sizes and dipole field strength for some selected extrasolar planets. The convective flux and velocity was calculated using an evolutionary age similar to that of our solar giants, 4.5 Gyr. The values shown for the magnetic moments are the average of the scaling models presented above, with the exception of Christensen et al. 2009.*

Inflated hot jupiters have been signaled out as plausible candidate for strong cyclotron radio emitters as a result of the interaction of their planetary magnetosphere with stellar wind. Assuming a radio emission mechanism similar to that of Jupiter, the flux output of this emission will be proportional to the magnetic/kinetic energy flux of the planetary magnetic field. The emitted characteristic cyclotron frequency will be set by the polar dipole field strength or the planetary magnetic moment through $f_c \sim 24\, M_E/R_p^3$ (MHz). Fig 6 shows our estimated values for the cyclotron frequencies for the planets considered here.

**Conclusions**

We have studied the dynamo action in 100 extrasolar giants in the current census based on a revised conductivity profile and a polytropic gas internal equation of state. Our new conductivity profile should be amenable to use in future MHD and magnetization studies of hydrogen-rich planets. The persistence of the convective action large enough to overcome ohmic dissipation and sustain the field generation was inferred from previous studies (Sanchez-Lavega 2004). We find that conducting cores in massive planets extend up to 0.98 of their planetary radii, implying that the poloidal dynamos in these planets might not be deep-seated. This is quite noteworthy especially in the light of the recent suggestion that atmospheric generated toroidal dynamos in sufficiently hot jupiters are also possible. Strong insolation should produce a steeper rise in internal conductivity profile and thus larger conducting core sizes than quoted in our analysis. Our estimates for closely-orbiting hot jupiters dynamo sizes should thus be regarded as lower bounds. We employed different scaling analyses to estimate the strength of the planetary magnetic field as well as their dipolar magnetic moments. For the most characterized planets, dipole surface fields of ~0.2-1.2 that of Jupiter or ~1.4-8.4 G are expected. Weaker dynamo fields should provide less protective shielding for upper atmospheric erosion arising from non-thermal ion-loss processes, and thus larger hydrogen mass loss rates. We find that the magnetic moments of extrasolar giants scale with $M_P R_p^3$, by a quadratic fit, consistent with our four solar giants.

Our predicted magnetic moments are in qualitative agreement with (Grießmeier, et al. 2004; Sanchez-Lavega 2004; Stevens 2005) studies. However, they are remarkably low compared to the most recent estimates based on (Reiners & Christensen 2010; Yadav & Thorngren 2017) scaling law, highlighting the role of reduced rotation rates in depressing dynamo strengths in all other scaling laws. In the latter results, the magnetic moment was taken to vary as the rate of heat flux, internal and incident, and thus the predicted fields are in the order of 100-300 G for planets with solar-like ages. However, for younger planets, with higher convective fluxes, fields of 1000 G were expected. The variance, by almost 2-3 orders of magnitude, is large enough that it would lead to dramatically different results regarding the magnetization of atmospheric flows, internal ohmic dissipation and thus the circulation & thermal history of these extrasolars. In this respect, the dispersion in estimates presents a marked opportunity for observational searches to constrain our fundamental knowledge of the dynamo action theory. Spectroscopic searches for Zeeman-splitted atomic transition lines of ionized species in the exospheres is already emerging as a favorable direction. This method is well established in stellar magnetic characterization and promises a direct probe to the field strength, since the magnitude of line splitting strength $\Delta \lambda_B \propto \lambda_0^2 B$, where $\lambda_0$ is the line central wavelength at zero field (Morin 2012). To date, the results found from the atomic linewidth of the H Lyman-$\alpha$ in the exosphere HD 209458b is quite promising and in agreement with our reported estimates. Our results demonstrate that the possible detection of exoplanetary radio emission above the 10-20 MHz ionospheric cutoff remains possible, albeit for a smaller target list than previous studies. The stronger emitters are expected to have maximal cyclotron frequencies between 20-50 MHz. This band should be detectable with the Low-Frequency Array-low band Antenna (LOFAR-LBA) and the proposed upgrade of the Very Large Array for planets at distances of several tens of parsecs.

Acknowledgments

The authors would like to thank Dave Stevenson for multiple insightful discussions on the Juno magnetic data. This material is based upon work supported by the Department of Energy National Nuclear Security Administration under Award Number DE-NA0001944, the University of Rochester, and the New York State Energy Research and Development Authority. The NASA Earth and Space Science Fellowship Program, Award NNX14AP17H, also supported this research.

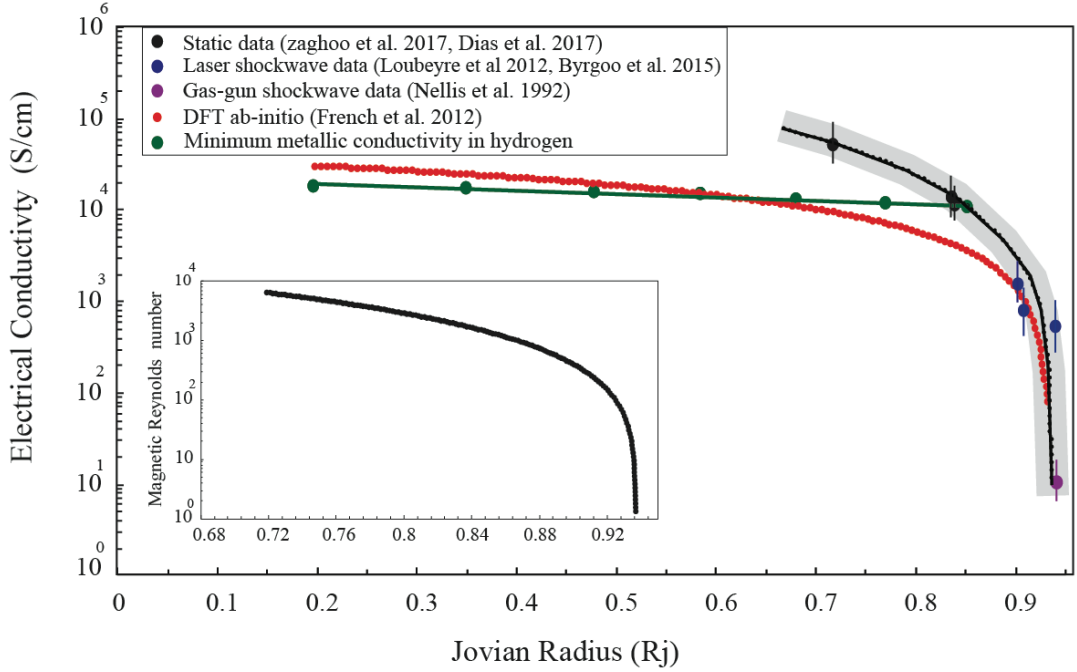

*FIG. 1. Electrical conductivity profile of pure hydrogen as a function of Jovian radius. The black circles refer to recent static laboratory experiments with their associated uncertainty (shaded grey area), while the green points are laser-driven shockwave experiments. The violet datum at $R_J=0.935$ is an earlier shockwave resistivity measurement; the black line is a quadratic fit to the data. All experimental data points, except for the highest pressure one, correspond to temperatures comparable to the Jovian adiabat at their respective pressures ~2000-5000 K. The dark green line represents the minimum metallic conductivity calculated for metallic hydrogen, whereas the red dashed is the theoretical ab-initio calculation along the adiabat. The inset shows the corresponding radial profile of the magnetic Reynolds number determined from the conductivity and the flow convective velocity.*

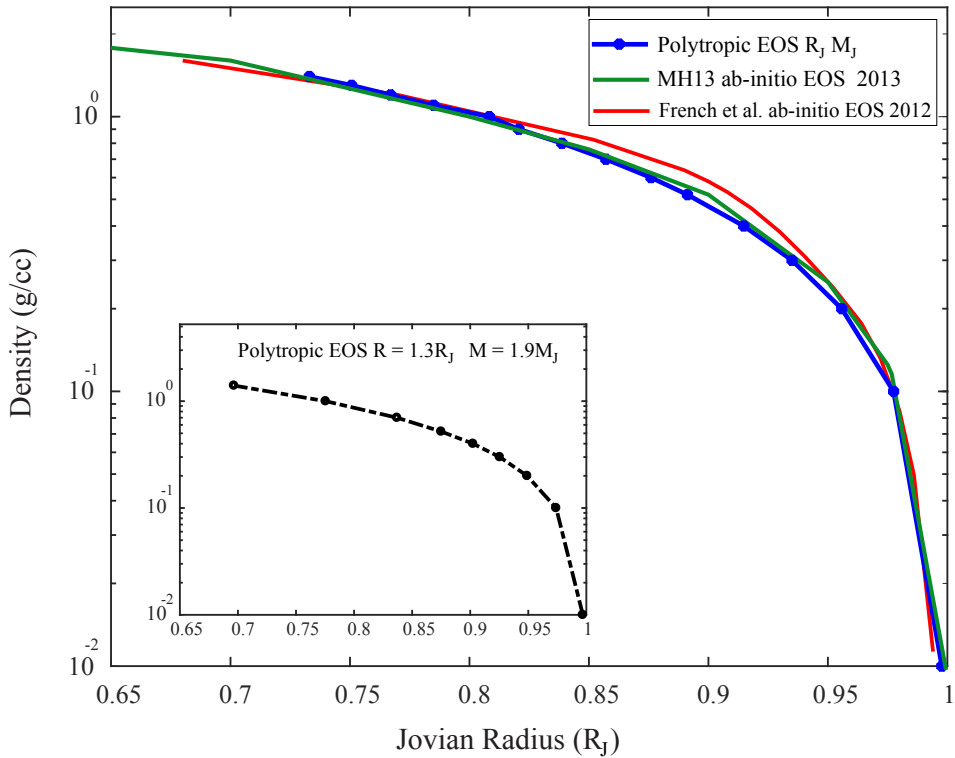

*FIG. 2 Internal density radial profile shown for different approximation: (blue line with circles) a polytropic equation of state with n=2, (Red line) the Militzer-Hubbard 2013 DFT-based EOS (Militzer & Hubbard 2013), (Green line) French et al. 2012 DFT-based EOS(French, et al. 2012). In the region between 0.99-0.95 Rj, the overlap is almost identical, however at lower depths, 0.95-0.85 Rj, the dispersion in the EOS is ~15%. This dispersion, though important for mass-radius relations and heavy elements distribution, have minimal effect in estimates for dynamo core sizes. The inset shows the calculated EOS for a Jupiter-like planet with a 1.9 Mj and 1.3 Rj.*

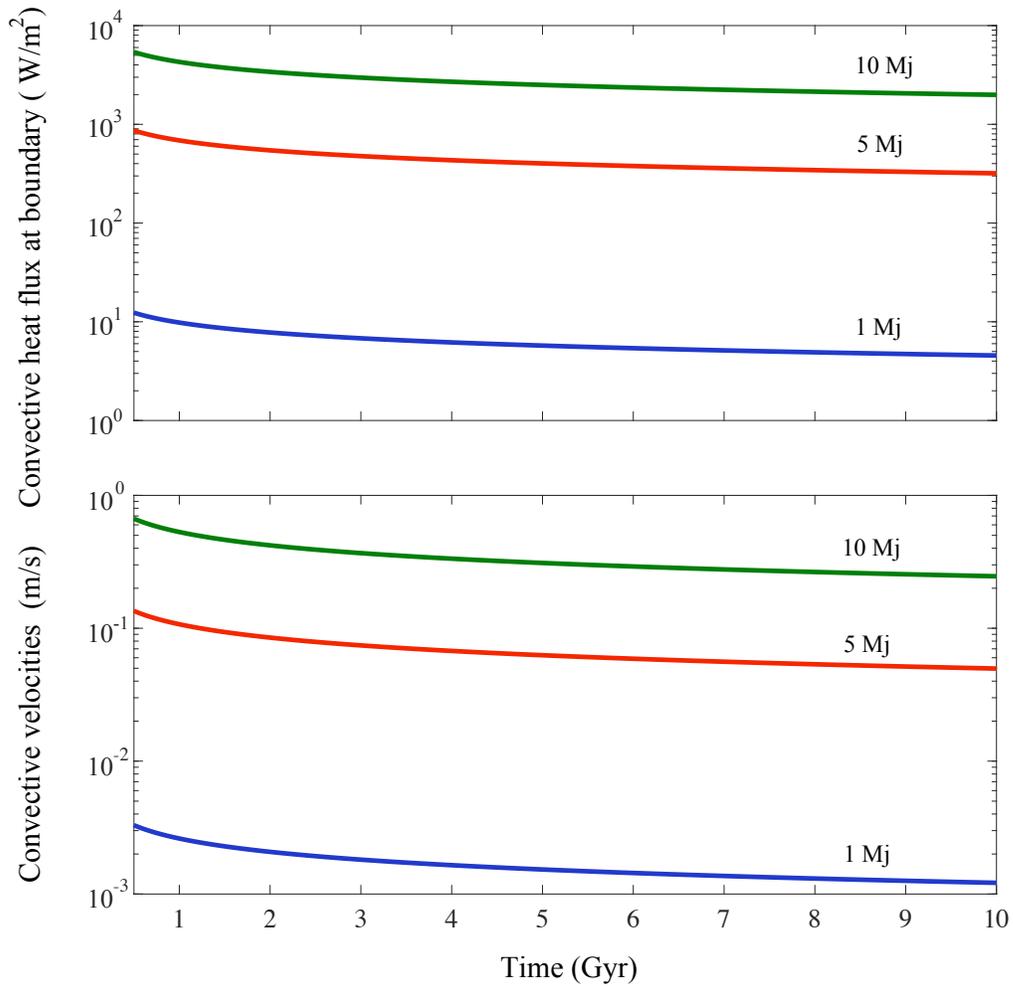

*FIG 3. Evolution of convective flow velocities (lower panel) and convective heat flux at the planetary surface boundary (top panel) for 3 selected Jupiter-like planets of masses 1,5,10 Mj and a constant radius of R=1.2Rj.*

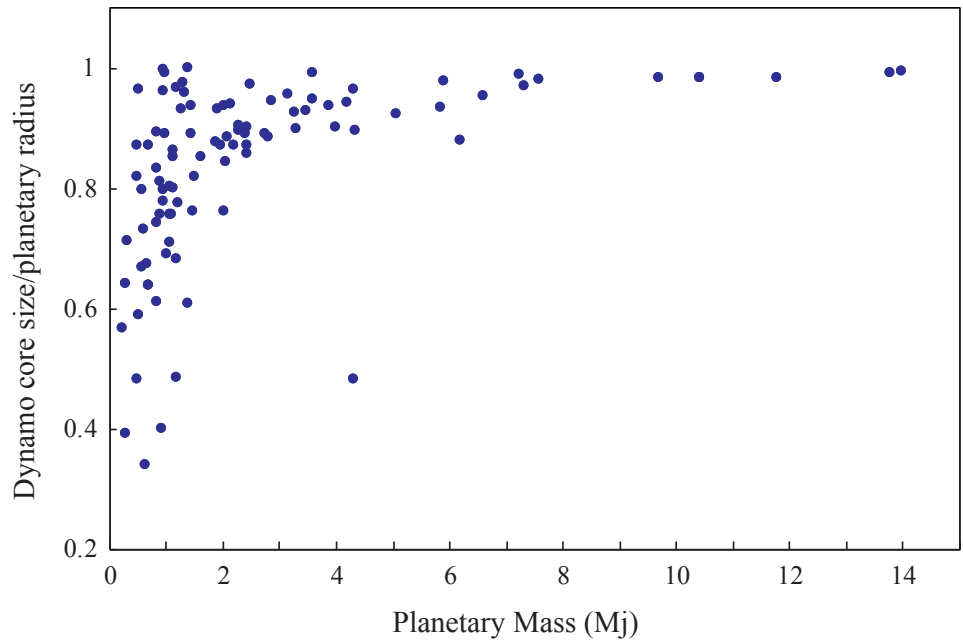

*FIG 4. Size of the dynamo conducting cores for extrasolar giants with respect to their respective planetary radius plotted as a function of their corresponding masses. We note that the size of the dynamo is not always increasing for more massive planets, as previously assumed in multiple scaling laws.*

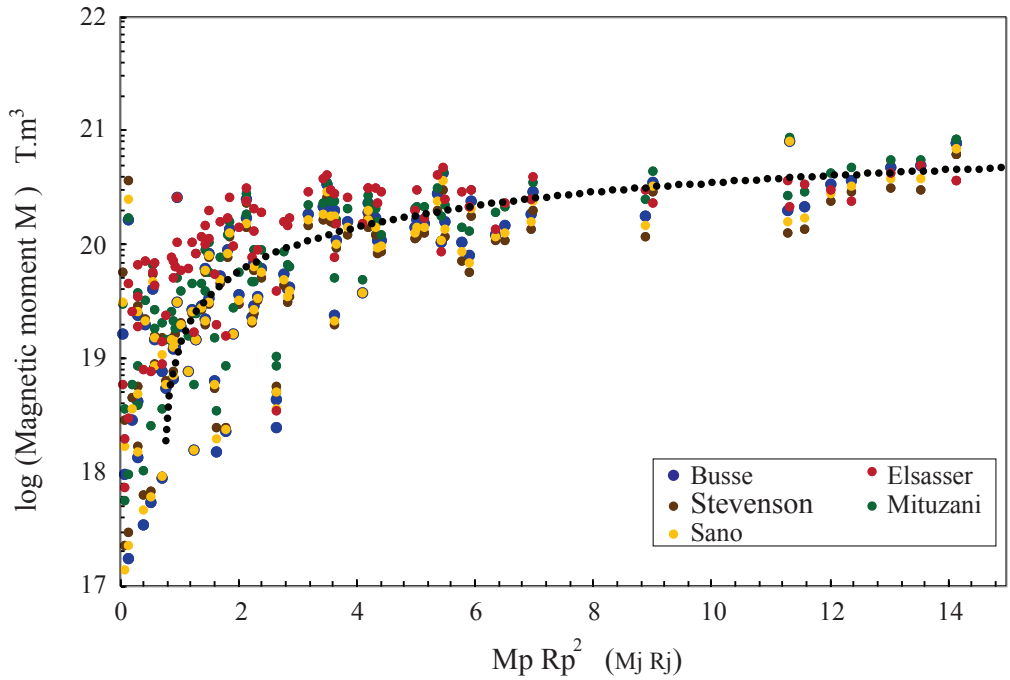

*FIG 5. Calculated magnetic moments for extrasolar giants plotted against the product of their respective mass and planetary radius squared, for different scaling laws shown in the legend. The black dotted line shows the quadratic fit to all the data. Masses and radii were scaled to the Jovian values. Notice that the range of calculated values appears insensitive to the choice of the scaling model.*

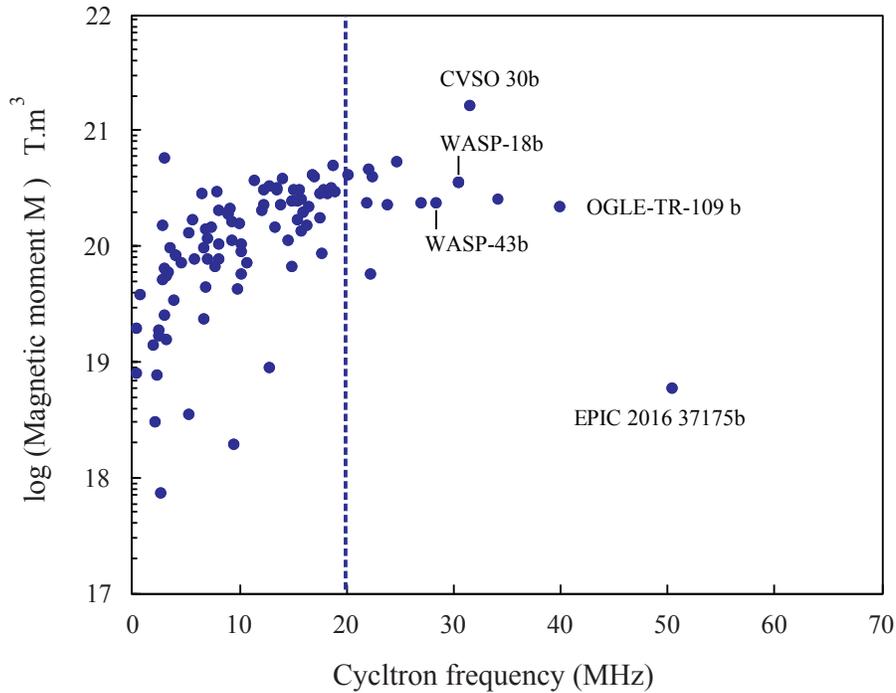

*FIG 6. Radio cyclotron frequencies as a function of the expected magnetic moments for the extrasolars giants considered in this study. The vertical line at 20 MHz is the upper bound for the Earth-ionospheric cutoff. Several strong emitters above this threshold are identified as promising candidates for future observational searches.*

Baraffe , I., Chabrier, G., Barman, T. S., Allard, F., & Hauschildt, P. H. 2003, Astronomy & Astrophysics, 402
Bastian, T. S., Dulk, G. A., & Leblanc, Y. 2000, ApJ, 545
Batygin, K., & Stevenson, D. J. 2010, ApJL, 714, L238
Bolton, S. J., et al. 2017, Science, 356, 821
Brygoo, S., et al. 2015, JAP, 118
Burrows, A., Hubbard, W. B., Lunine, J. I., & Liebert, J. 2001, Rev Mod Phys, 73, 719
Burrows, A., et al. 1997, ApJ, 491, 856
Busse, F. H. 1976, Phys Earth Planet Inter, 12
Christensen, U. R. 2010, Space Sci Rev, 152, 565
Christensen, U. R., & Aubert, J. 2006, Geophys J Int, 166, 97
Connerney, J. E. P., et al. 2017, Science, 356, 826
Cuntz, M., Saar, S. H., & Musielak, Z. E. 2000, ApJ, 533
Curtis, S. A., & Ness, N. F. 1986, J Geophys Res, 91, 11003
Dias, R. P., & Silvera, I. F. 2017, Science, 355, 715
Duarte, L. D., Wicht, J., & Gastine, T. 2018, Icarus, 299, 206
Farrell, W. M., Desch, M. D., & Zarka, P. 1999, J Geophys Res, 104, 14025
French, M., Becker, A., Lorenzen, W., Nettelmann, N., Bethkenhagen, M., Wicht, J., & Redmer, R. 2012, Astrophys J Suppl, 202

Grießmeier, J. M. 2015, Astrophysics and Space Science Library, 411, 213
Grießmeier, J. M., Mostchmann, U., Mann, G., & Rucker, H. O. 2005, Astronomy & Astrophysics, 437, 717
Grießmeier, J. M., et al. 2004, Astronomy & Astrophysics, 425, 753
Hubbard, W. B. 1977, Icarus, 30, 305
Hubbard, W. B., Guillot, T., & Lunine, J. I. 1997, Phys of Plasmas, 4, 2011
Hubbard, W. B., & Stevenson, D. J. 1984, Saturn, Univ of Arizona Press, 47
Joseph, T., et al. 2004, ApJ, 612
Kislyakova, K., Holmstrom, M., Lammer, H., Odert, P., & Khodachenko, M. 2014, Science, 346
Menou, K. 2012, ApJ, 745, 138
Militzer, B., & Hubbard, W. B. 2013, ApJ, 774
Mizutani, H., Yamamato, T., & Fujimura, A. 1992, Adv in Space Research, 12, 265
Morin, J. 2012, EAS Publ Ser, 57, 165
Mott, N. F. 1972, Phil Mag, 26
Nellis, W. J., Mitchell, A. C., McCandless, P. C., Erskine, D. J., & Weir, S. T. 1992, Phys Rev Lett, 68, 2937
Perna, R., Menou, K., & Rauscher, E. 2010, ApJ, 724
Reiners, A., & Christensen, U. R. 2010, Astronomy & Astrophysics, 522
Rogers, T. M. 2017, Nature Astronomy, 1
Sanchez-Lavega, A. 2004, ApJ, 609, 87
Sano, Y. 1993, Geomag Geoelectr 45, 65
Showman, A., & Guillot, T. 2002, Astronomy & Astrophysics, 385, 166
Sirothia, S. K., Lecavelier des Etangs, A., Gopal-Krishna, Kantharia, N. G., & Ishwar-Chandra, C. H. 2014, Astronomy & Astrophysics, 562
Stevens, I. R. 2005, MNRAS, 356, 1053
Stevenson, D. J. 1983, Rep Prog Phys, 46, 555
Trammell, G. B., Arras, P., & Li, Z.-Y. 2011, ApJ, 728
Yadav, R. K., & Thorngren, D. P. 2017, ApJL, 849
Zaghoo, M., Salamat, A., & Silvera, I. F. 2016, Phys Rev B, 93, 155128
Zaghoo, M., & Silvera, I. F. 2017, PNAS, 10